\documentclass[a4paper,12pt]{revtex4}
\usepackage[utf8]{inputenc}
\usepackage[T1]{fontenc}
\usepackage{lmodern}
\usepackage{epsfig}
\usepackage{array}
\usepackage{amsmath}
\usepackage{epstopdf}
\usepackage{graphicx}
\graphicspath{/}
\begin{document}

\title{Definite evidence for physically incomprehensible inequality $\sigma_{tot}(e^+e^- \to n \bar n) > \sigma_{tot}(e^+e^- \to p \bar p)$}.
\date{\today}

\medskip

\author{Stanislav Dubni\v cka}
\address{Institute of Physics, Slovak Academy of Sciences,
Bratislava, Slovak Republic}

\author{Anna Z. Dubni\v ckov\'a}
\address{Department of Theoretical Physics, Comenius University,
Bratislava, Slovak Republic}

\begin{abstract}
   Though the neutron mass is larger than the proton mass and obviously one expects in $e^+e^-$ annihilation
a creation of more $p \bar p$ pairs in comparison with $n \bar n$ at the same energy, just the opposite inequality for the corresponding
total cross sections $\sigma_{tot}(e^+e^- \to n \bar n) > \sigma_{tot}(e^+e^- \to p \bar p)$ has been revealed in the analysis
of only the present time existing proton electromagnetic form factors data by the unitary and analytic approach.
\end{abstract}


\maketitle

\section{Introduction}

   From all particles in \cite{PDG} only the protons $p$ and electrons $e$ and their antiparticles antiprotons $\bar p$ and positrons $e^+$ are stable. While the
electrons and positrons in interactions with other particles behave as point-like up to now, the protons and antiprotons are compound of three quarks and
in interaction with electrons and positrons they behave as extended objects in the space. This property of protons and antiprotons is called the electromagnetic (EM) structure, for the first time revealed for the proton in the elastic scattering process $e^-p \to e^-p$ at the middle of the last century, before the quark model of strongly interacting particles, hadrons, has been established. Later on this non-point-like nature of the proton has been generalized also to all other hadrons, including the antiproton and the isotopic partner of the proton, the unstable neutron.

   For a complete description of the EM structure of the proton and also neutron one needs two scalar functions, to be called form
factors (FFs), in the spacelike region depending on the momentum transfer squared $t=-Q^2$ and in the timelike region on the squared total c.m.
energy $s=w^2$. In the choice of the forms of FFs there is some freedom.

   To know EM structure of protons and neutrons means to have the adequate multitude of as precise as possible experimental information on their FFs in the
whole region of their definition and to dispose with reliable theoretical models to be able predict FFs behaviors consistently with data.

   Because more precise experimental data (though far off to be ideal) exist on the proton form factors than on the form
factors of the neutron, mainly due to the stability of the proton, we have predicted in our paper \cite{DD} the neutron EM FFs behaviors in the whole region of their
definition correctly theoretically. This has been realized by using the advanced neutron EM structure $Unitary\&Analytic$ model \cite{ABDD}, the parameters of which are numerically evaluated in a description of the proton EM FFs data by the advanced proton EM structure $Unitary\&Analytic$ model \cite{ABDD} and
an agreement of existing neutron data with such theoretical prediction is amazing. In this way behaviors of both total cross sections, $\sigma_{tot}(e^+e^- \to p \bar p)$ and $\sigma_{tot}(e^+e^- \to n \bar n)$, are known and by their comparison in one figure the physically incomprehensible inequality $\sigma_{tot}(e^+e^- \to n \bar n) > \sigma_{tot}(e^+e^- \to p \bar p)$ has been revealed, awaiting for experimental confirmation.

\section{Proton and neutron electromagnetic structure models}

   The electromagnetic (EM) structure of the proton and neutron can be completely described theoretically e.g. by the Dirac $F^p_1(t),F^n_1(t)$ and Pauli $F^p_2(t),F^n_2(t)$
FFs, which naturally appear in a decomposition of the proton
and neutron matrix elements of the EM current $J^{EM}_\mu (0)$ as coefficients of only two
linearly independent covariants constructed from the four momenta $p, p'$,
$\gamma$-matrices and Dirac bi-spinors
\begin{small}
\begin{eqnarray}
  <p|J^{EM}_\mu (0)|p>=e \bar u(p')[\gamma_\mu F^p_1(t)+\frac{i}{2m_p}
  \sigma_{\mu \nu}(p'-p)_\mu F^p_2(t)] u(p),
\end{eqnarray}
\end{small}
and
\begin{small}
\begin{eqnarray}
  <n|J^{EM}_\mu (0)|n>=e \bar u(p')[\gamma_\mu F^n_1(t)+\frac{i}{2m_n}
  \sigma_{\mu \nu}(p'-p)_\mu F^n_2(t)] u(p),
\end{eqnarray}
\end{small}
with $m_p$ and $m_n$ to be the proton and neutron mass, respectively.

   A description of the proton and neutron EM structures is even improved if mixed transformation properties of the
EM current $J^{EM}_\mu (0)$ under the rotation in the isospin space is utilized. A part of $J^{EM}_\mu (0)$ transforms as an isoscalar and its another part as the third component of the isovector. The latter leads to a splitting of the proton and neutron Dirac and Pauli FFs into the same flavor-independent isoscalar and isovector parts as follows
\begin{small}
\begin{eqnarray}
 F^p_1(t)=[F^N_{1s}(t)+F^N_{1v}(t)],\nonumber\\
 F^p_2(t)=[F^N_{2s}(t)+F^N_{2v}(t)],\label{DPp}
\end{eqnarray}
\end{small}
\begin{small}
\begin{eqnarray}
 F^n_1(t)=[F^N_{1s}(t)-F^N_{1v}(t)],\nonumber\\
 F^n_2(t)=[F^N_{2s}(t)-F^N_{2v}(t)],\label{DPn}
\end{eqnarray}
\end{small}
whereby, the sign between them is specified by the sign of the third component of the
proton and neutron isospin, respectively.

   The FFs $F^N_{1s}(t), F^N_{1v}(t), F^N_{2s}(t), F^N_{2v}(t)$ are analytic in the whole complex t-plane
besides the cuts on the positive real axis starting from the lowest possible branch points of them. Moreover they
contain all known properties of the proton and neutron FFs, like experimental fact of a production of
unstable vector-meson resonances in $e^+e^-$ annihilation  processes into hadrons, normalization conditions,
the asymptotic behaviors as predicted by the quark model of hadrons and fulfil reality conditions and
unitarity conditions.

   In the paper \cite{ABDD} just such advanced 9 vector-meson resonance\\
 $\rho(770), \omega(782), \phi(1020); \rho'(1450), \omega'(1420), \phi'(1680); \rho''(1700), \omega''(1650), \phi''(2170);$  \cite{PDG}\\
Unitary and Analytic $(U\&A)$ model for the proton and neutron isoscalar and isovector Dirac and Pauli FFs has been constructed
\begin{eqnarray}\label{FN1s}\nonumber
  F^N_{1s}[V(t)]=\Bigg(\frac{1-V^2}{1-V^2_N}\Bigg)^4\Bigg\{\frac{1}{2}H_{\omega''}(V)H_{\phi''}(V)\\\nonumber
  +\Bigg[H_{\phi''}(V)H_{\omega'}(V)\frac{(C^{1s}_{\phi''}-C^{1s}_{\omega'})}{(C^{1s}_{\phi''}-C^{1s}_{\omega''})}+
  H_{\omega''}(V)H_{\omega'}(V)\frac{(C^{1s}_{\omega''}-C^{1s}_{\omega'})}{(C^{1s}_{\omega''}-C^{1s}_{\phi''})}\\
  -H_{\omega''}(V)H_{\phi''}(V)\Bigg](f^{(1)}_{\omega'NN}/f_{\omega'})\nonumber\\
  +\Bigg[H_{\phi''}(V)H_{\phi'}(V)\frac{(C^{1s}_{\phi''}-C^{1s}_{\phi'})}{(C^{1s}_{\phi''}-C^{1s}_{\omega''})}+
  H_{\omega''}(V)H_{\phi'}(V)\frac{(C^{1s}_{\omega''}-C^{1s}_{\phi'})}{(C^{1s}_{\omega''}-C^{1s}_{\phi''})}\nonumber\\
  -H_{\omega''}(V)H_{\phi''}(V)\Bigg](f^{(1)}_{\phi'NN}/f_{\phi'})\\\nonumber
  +\Bigg[H_{\phi''}(V)L_{\omega}(V)\frac{(C^{1s}_{\phi''}-C^{1s}_{\omega})}{(C^{1s}_{\phi''}-C^{1s}_{\omega''})}+
  H_{\omega''}(V)L_{\omega}(V)\frac{(C^{1s}_{\omega''}-C^{1s}_{\omega})}{(C^{1s}_{\omega''}-C^{1s}_{\phi''})}\\\nonumber
  -H_{\omega''}(V)H_{\phi''}(V)\Bigg](f^{(1)}_{\omega NN}/f_{\omega})\\\nonumber
  +\Bigg[H_{\phi''}(V)L_{\phi}(V)\frac{(C^{1s}_{\phi''}-C^{1s}_{\phi})}{(C^{1s}_{\phi''}-C^{1s}_{\omega''})}+
  H_{\omega''}(V)L_{\phi}(V)\frac{(C^{1s}_{\omega''}-C^{1s}_{\phi})}{(C^{1s}_{\omega''}-C^{1s}_{\phi''})}\\
  -H_{\omega''}(V)H_{\phi''}(V)\Bigg](f^{(1)}_{\phi NN}/f_{\phi})\Bigg\}\nonumber
\end{eqnarray}
with 5 free parameters
$(f^{(1)}_{\omega'NN}/f_{\omega'}), (f^{(1)}_{\phi'NN}/f_{\phi'}),
(f^{(1)}_{\omega NN}/f_{\omega}), (f^{(1)}_{\phi NN}/f_{\phi}),
t^{1s}_{in}$
\begin{eqnarray}\label{FN1v}\nonumber
  F^N_{1v}[W(t)]=\Bigg(\frac{1-W^2}{1-W^2_N}\Bigg)^4\Bigg\{\frac{1}{2}L_\rho(W)L_{\rho'}(W)\\
  +\Bigg[L_{\rho'}(W)L_{\rho''}(W)\frac{(C^{1v}_{\rho'}-C^{1v}_{\rho''})}{(C^{1v}_{\rho'}-C^{1v}_\rho)}+
  L_\rho(W)L_{\rho''}(W)\frac{(C^{1v}_\rho-C^{1v}_{\rho''})}{(C^{1v}_\rho-C^{1v}_{\rho'})}\\\nonumber
  -L_\rho(W)L_{\rho'}(W)\Bigg](f^{(1)}_{\rho NN}/f_{\rho})\Bigg\}
\end{eqnarray}
with 2 free parameters
$(f^{(1)}_{\rho NN}/f_{\rho})$ and $t^{1v}_{in}$,
\begin{eqnarray}\label{FN2s}\nonumber
  F^N_{2s}[U(t)]=\Bigg(\frac{1-U^2}{1-U^2_N}\Bigg)^6\Bigg\{\frac{1}{2}(\mu_p+\mu_n-1)H_{\omega''}(U)H_{\phi''}(U)H_{\omega'}(U)\\\nonumber
  +\Bigg[H_{\phi''}(U)H_{\omega'}(U)H_{\phi'}(U)\frac{(C^{2s}_{\phi''}-C^{2s}_{\phi'})(C^{2s}_{\omega'}-C^{2s}_{\phi'})}
  {(C^{2s}_{\phi''}-C^{2s}_{\omega''})(C^{2s}_{\omega'}-C^{2s}_{\omega''})}\\\nonumber
  +H_{\omega''}(U)H_{\omega'}(U)H_{\phi'}(U)\frac{(C^{2s}_{\omega''}-C^{2s}_{\phi'})(C^{2s}_{\omega'}-C^{2s}_{\phi'})}
  {(C^{2s}_{\omega''}-C^{2s}_{\phi''})(C^{2s}_{\omega'}-C^{2s}_{\phi''})}\\\nonumber
  +H_{\omega''}(U)H_{\phi''}(U)H_{\phi'}(U)\frac{(C^{2s}_{\omega''}-C^{2s}_{\phi'})(C^{2s}_{\phi''}-C^{2s}_{\phi'})}
  {(C^{2s}_{\omega''}-C^{2s}_{\omega'})(C^{2s}_{\phi''}-C^{2s}_{\omega'})}\\\nonumber
  -H_{\omega''}(U)H_{\phi''}(U)H_{\omega'}(U)\Bigg](f^{(2)}_{\phi'NN}/f_{\phi'})\\\nonumber
  +\Bigg[H_{\phi''}(U)H_{\omega'}(U)L_{\omega}(U)\frac{(C^{2s}_{\phi''}-C^{2s}_{\omega})(C^{2s}_{\omega'}-C^{2s}_{\omega})}
  {(C^{2s}_{\phi''}-C^{2s}_{\omega''})(C^{2s}_{\omega'}-C^{2s}_{\omega''})}\\\nonumber
  +H_{\omega''}(U)H_{\omega'}(U)L_{\omega}(U)\frac{(C^{2s}_{\omega''}-C^{2s}_{\omega})(C^{2s}_{\omega'}-C^{2s}_{\omega})}
  {(C^{2s}_{\omega''}-C^{2s}_{\phi''})(C^{2s}_{\omega'}-C^{2s}_{\phi''})}+\\
  +H_{\omega''}(U)H_{\phi''}(U)L_{\omega}(U)\frac{(C^{2s}_{\omega''}-C^{2s}_{\omega})(C^{2s}_{\phi'}-C^{2s}_{\omega})}
  {(C^{2s}_{\omega''}-C^{2s}_{\omega'})(C^{2s}_{\phi''}-C^{2s}_{\omega'})}\\\nonumber
  -H_{\omega''}(U)H_{\phi''}(U)H_{\omega'}(U)\Bigg](f^{(2)}_{\omega NN}/f_{\omega})\\\nonumber
  +\Bigg[H_{\phi''}(U)H_{\omega'}(U)L_{\phi}(U)\frac{(C^{2s}_{\phi''}-C^{2s}_{\phi})(C^{2s}_{\omega'}-C^{2s}_{\phi})}
  {(C^{2s}_{\phi''}-C^{2s}_{\omega''})(C^{2s}_{\omega'}-C^{2s}_{\omega''})}\\\nonumber
  +H_{\omega''}(U)H_{\omega'}(U)L_{\phi}(U)\frac{(C^{2s}_{\omega''}-C^{2s}_{\phi})(C^{2s}_{\omega'}-C^{2s}_{\phi})}
  {(C^{2s}_{\omega''}-C^{2s}_{\phi''})(C^{2s}_{\omega'}-C^{2s}_{\phi''})}\\\nonumber
  +H_{\omega''}(U)H_{\phi''}(U)L_{\phi}(U)\frac{(C^{2s}_{\omega''}-C^{2s}_{\phi})(C^{2s}_{\phi''}-C^{2s}_{\phi})}
  {(C^{2s}_{\omega''}-C^{2s}_{\omega'})(C^{2s}_{\phi''}-C^{2s}_{\omega'})}\\\nonumber
  -H_{\omega''}(U)H_{\phi''}(U)H_{\omega'}(U)\Bigg](f^{(2)}_{\phi NN}/f_{\phi})\Bigg\}
\end{eqnarray}
with 4 free parameters
$(f^{(2)}_{\phi'NN}/f_{\phi'})$, $(f^{(2)}_{\omega NN}/f_{\omega})$,
$(f^{(2)}_{\phi NN}/f_{\phi}), t^{2s}_{in}$, and
\begin{eqnarray}\label{FN2v}
  F^N_{2v}[X(t)]=\Bigg(\frac{1-X^2}{1-X^2_N}\Bigg)^6\Bigg\{\frac{1}{2}(\mu_p-\mu_n-1)L_\rho(X)L_{\rho'}(X)H_{\rho''}(X)\Bigg\}
\end{eqnarray}
to be dependent on only 1 free parameter $t^{2v}_{in}$, where\\
$V(t)=i\frac{\sqrt{(\frac{t^{1s}_{in}-t^{1s}_0}{t^{1s}_0})^{1/2}+(\frac{t-t^{1s}_0}{t^{1s}_0})^{1/2}}-\sqrt{(\frac{t^{1s}_{in}-t^{1s}_0}{t^{1s}_0})^{1/2}-(\frac{t-t^{1s}_0}{t^{1s}_0})^{1/2}}}
            {\sqrt{(\frac{t^{1s}_{in}-t^{1s}_0}{t^{1s}_0})^{1/2}+(\frac{t-t^{1s}_0}{t^{1s}_0})^{1/2}}+\sqrt{(\frac{t^{1s}_{in}-t^{1s}_0}{t^{1s}_0})^{1/2}-(\frac{t-t^{1s}_0}{t^{1s}_0})^{1/2}}},$
similarly  $W(t), U(t), X(t)$, are conformal mappings of the corresponding four-sheeted Riemann surfaces in $t$ variable, on which the corresponding FFs are defined,
always into one $V-, W-, U-, X-$ plane. The $t^{1s}_0=9m_\pi^2$, $t^{1v}_0=4m_\pi^2$, $t^{2s}_0=9m_\pi^2$,  $t^{2v}_0=4m_\pi^2$ are the lowest branch points of FFs and
$t^{1s}_{in}$, $t^{1v}_{in}$, $t^{2s}_{in}$,  $t^{2v}_{in}$ are the effective inelastic square root branch points, representing contributions of all possible higher inelastic thresholds effectively and therefore they are left in the analysis of data as free parameters and $V_N=V(0)$, similarly $W_N, U_N, X_N$, are the positions in those planes corresponding to normalizations of FFs at t=0.

   Denotations $L$ (lower) and $H$ (higher)
\begin{eqnarray}
  L_r(V)=\frac{(V_N-V_r)(V_N-V^*_r)(V_N-1/V_r)(V_N-1/V^*_r)}{(V-V_r)(V-V^*_r)(V-1/V_r)(V-1/V^*_r)},\\\label{eq19}
  C^{1s}_r=\frac{(V_N-V_r)(V_N-V^*_r)(V_N-1/V_r)(V_N-1/V^*_r)}{-(V_r-1/V_r)(V_r-1/V^*_r)}, r=\omega, \phi \nonumber
\end{eqnarray}
\begin{eqnarray}
  H_l(V)=\frac{(V_N-V_l)(V_N-V^*_l)(V_N+V_l)(V_N+V^*_l)}{(V-V_l)(V-V^*_l)(V+V_l)(V+V^*_l)},\\\label{eq20}
  C^{1s}_l=\frac{(V_N-V_l)(V_N-V^*_l)(V_N+V_l)(V_N+V^*_l)}{-(V_l-1/V_l)(V_l-1/V^*_l)}, l=
  \omega'', \phi'', \omega', \phi' \nonumber
\end{eqnarray}
\begin{eqnarray}
  L_k(W)=\frac{(W_N-W_k)(W_N-W^*_k)(W_N-1/W_k)(W_N-1/W^*_k)}{(W-W_k)(W-W^*_k)(W-1/W_k)(W-1/W^*_k)},\\ \label{eq21}
  C^{1v}_k=\frac{(W_N-W_k)(W_N-W^*_k)(W_N-1/W_k)(W_N-1/W^*_k)}{-(W_k-1/W_k)(W_k-1/W^*_k)}, k=\rho'',
  \rho', \rho \nonumber
\end{eqnarray}
\begin{eqnarray}
  L_r(U)=\frac{(U_N-U_r)(U_N-U^*_r)(U_N-1/U_r)(U_N-1/U^*_r)}{(U-U_r)(U-U^*_r)(U-1/U_r)(U-1/U^*_r)},\\ \label{eq22}
  C^{2s}_r=\frac{(U_N-U_r)(U_N-U^*_r)(U_N-1/U_r)(U_N-1/U^*_r)}{-(U_r-1/U_r)(U_r-1/U^*_r)}, r=\omega, \phi \nonumber
\end{eqnarray}
\begin{eqnarray}
  H_l(U)=\frac{(U_N-U_l)(U_N-U^*_l)(U_N+U_l)(U_N+U^*_l)}{(U-U_l)(U-U^*_l)(U+U_l)(U+U^*_l)},\\ \label{eq23}
  C^{2s}_l=\frac{(U_N-U_l)(U_N-U^*_l)(U_N+U_l)(U_N+U^*_l)}{-(U_l-1/U_l)(U_l-1/U^*_l)}, l=
  \omega'', \phi'', \omega', \phi' \nonumber
\end{eqnarray}
\begin{eqnarray}
  L_k(X)=\frac{(X_N-X_k)(X_N-X^*_k)(X_N-1/X_k)(X_N-1/X^*_k)}{(X-X_k)(X-X^*_k)(X-1/X_k)(X-1/X^*_k)},\\\label{eq24}
  C^{2v}_k=\frac{(X_N-X_k)(X_N-X^*_k)(X_N-1/X_k)(X_N-1/X^*_k)}{-(X_k-1/X_k)(X_k-1/X^*_k)}, k=\rho', \rho \nonumber
\end{eqnarray}
\begin{eqnarray}
  H_{\rho''}(X)=\frac{(X_N-X_{\rho''})(X_N-X^*_{\rho''})(X_N+X_{\rho''})(X_N+X^*_{\rho''})}
  {(X-X_{\rho''})(X-X^*_{\rho''})(X+X_{\rho''})(X+X^*_{\rho''})},\\ \label{eq25}
  C^{2v}_{\rho''}=\frac{(X_N-X_{\rho''})(X_N-X^*_{\rho''})(X_N+X_{\rho''})(X_N+X^*_{\rho''})}
  {-(X_{\rho''}-1/X_{\rho''})(X_{\rho''}-1/X^*_{\rho''})}.\nonumber
\end{eqnarray}
mean that in the first case a real part of the resonance location in the complex t-plane of some unphysical sheet
of the four-sheeted Riemann surface is found below the corresponding effective
inelastic square root branch point and in the second case a real part of the resonance
location is found above the corresponding effective inelastic square root branch point.

   Whereas the Dirac $F^p_1(t)$ and Pauli $F^p_2(t)$ proton FFs, as it is seen from the previous, are very effective for theoretical considerations of the proton EM
structure, for extraction of an experimental information on it from measured cross sections and polarizations the proton electric $G^p_E(t)$ and proton magnetic $G^p_M(t)$ FFs are more comfortable. Their contributions e.g. into the total cross section of the $e^+e^- \to p \bar p$ process  \cite{BPZZ} are separated
\begin{eqnarray}\label{totcspp}
 \sigma_{tot}(e^+e^- \to p \bar p)=\frac{4 \pi \alpha^2 C \beta_p(t)}{3 t}
 [|G^p_M(t)|^2+\frac{2m_p}{t}|G^p_E(t)|^2],
\end{eqnarray}
contributions of the Dirac and Pauli FFs appear there with an interference term. In (\ref{totcspp}) $\beta_p(t)=\sqrt{1-\frac{4 m^2_p}{t}}$ and $C$
is the so-called Sommerfeld-Gamov-Sakharov enhancement factor.
   A similar expression for the total cross section of the neutrons can be written
   \begin{eqnarray}\label{totcsnn}
 \sigma_{tot}(e^+e^- \to n \bar n)=\frac{4 \pi \alpha^2 \beta_n(t)}{3 t}
 [|G^n_M(t)|^2+\frac{2m_n}{t}|G^n_E(t)|^2]
\end{eqnarray}
however, now without the Coulomb enhancement factor $C$, which in the case of the neutral final state particles is equal 1.

   The relations between both sets of FFs are

\begin{eqnarray}\label{pEMFFs}
  G^p_E(t)=[F^N_{1s}(t)+F^N_{1v}(t)]+
  \frac{t}{4 m^2_p}[F^N_{2s}(t)+F^N_{2v}(t)]\\
  G^p_M(t)=[F^N_{1s}(t)+F^N_{1v}(t)]+[F^N_{2s}(t)+F^N_{2v}(t)]\nonumber
\end{eqnarray}
and
\begin{eqnarray}\label{nEMFFs}
  G^n_E(t)=[F^N_{1s}(t)-F^N_{1v}(t)]+
  \frac{t}{4 m^2_n}[F^N_{2s}(t)-F^N_{2v}(t)]\\
  G^n_M(t)=[F^N_{1s}(t)-F^N_{1v}(t)]+[F^N_{2s}(t)-F^N_{2v}(t)]\nonumber
\end{eqnarray}
with normalizations
\begin{eqnarray}
  G^p_E(0)= 1;\quad G^p_M(0)=\mu_p;
\end{eqnarray}
\begin{eqnarray}
  G^n_E(0)= 0;\quad G^n_M(0)=\mu_n.
\end{eqnarray}
and
\begin{eqnarray}
  F^N_{1s}(0)=F^N_{1v}(0)=\frac{1}{2};\quad F^N_{2s}(0)=\frac{1}{2}(\mu_p+\mu_n-1);\quad F^N_{2v}(0)=\frac{1}{2}(\mu_p-\mu_n-1),
\end{eqnarray}
where $\mu_p$ and $\mu_n$ are the magnetic moments of the proton and neutron, respectively.

   The model is in fact a well-matched unification of pole contributions of unstable vector mesons with cut structure in the complex plane $t$,
whereby these cuts represent so-called continua contributions generated by exchange of more than one particle in the corresponding Feynman diagrams and they secure FFs imaginary parts to be different from zero just beyond the lowest possible thresholds on the positive real axis, as it is required by the FFs unitarity conditions.

   There are approximately 460 more or less reliable experimental points on the proton EM FFs $G^p_E(t), G^p_M(t)$ and their ratios \cite{jones}-\cite{Ab1}.
Results of their analysis by the proton EM structure $U\&A$ model given by (\ref{FN1s})-(\ref{FN2v}) and
(\ref{pEMFFs}) with 12 free parameters with clear physical meaning are given in TABLE I.

\begin{figure}
    \includegraphics[width=0.35\textwidth]{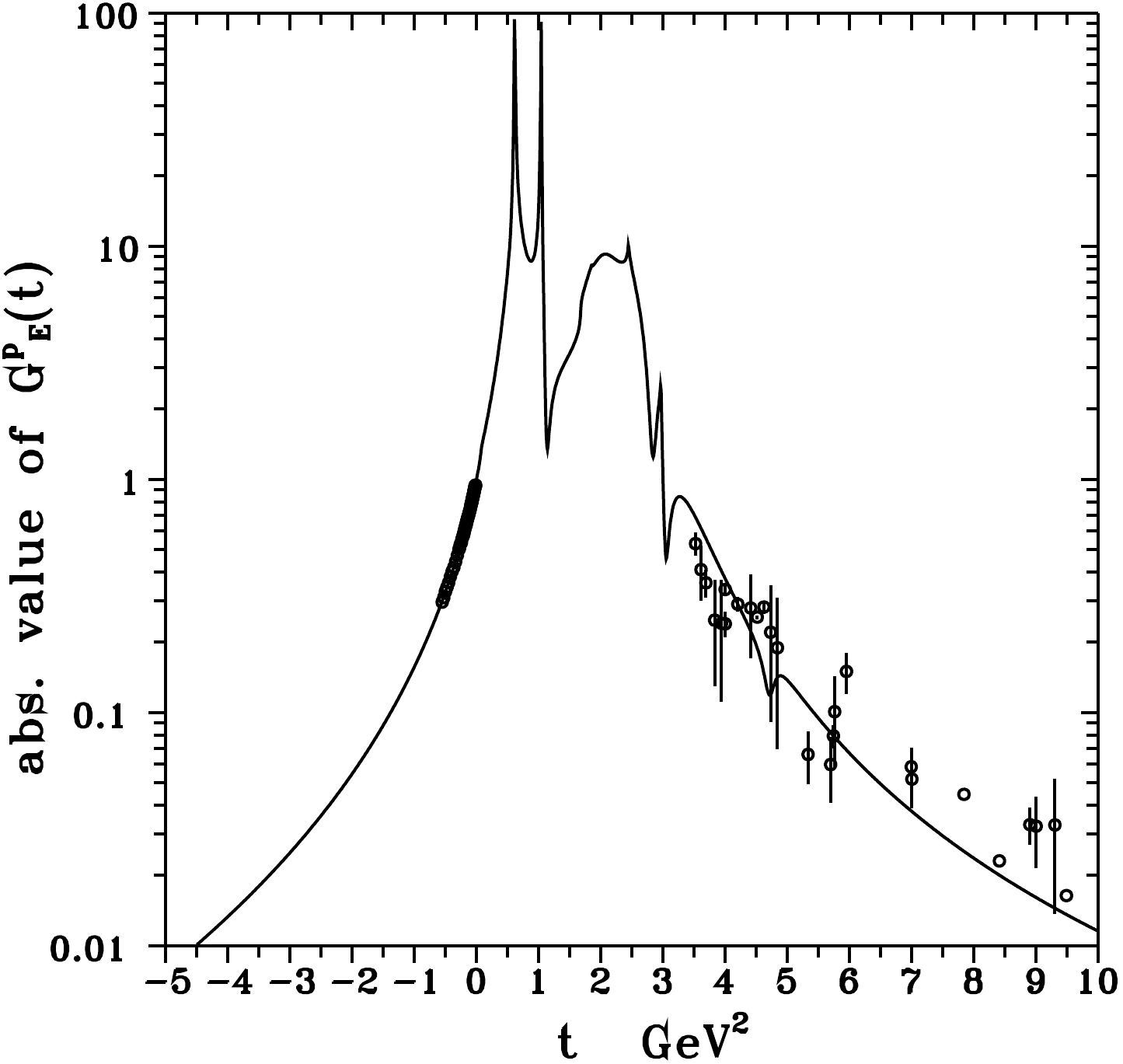}\hspace{0.3cm}
    \includegraphics[width=0.35\textwidth]{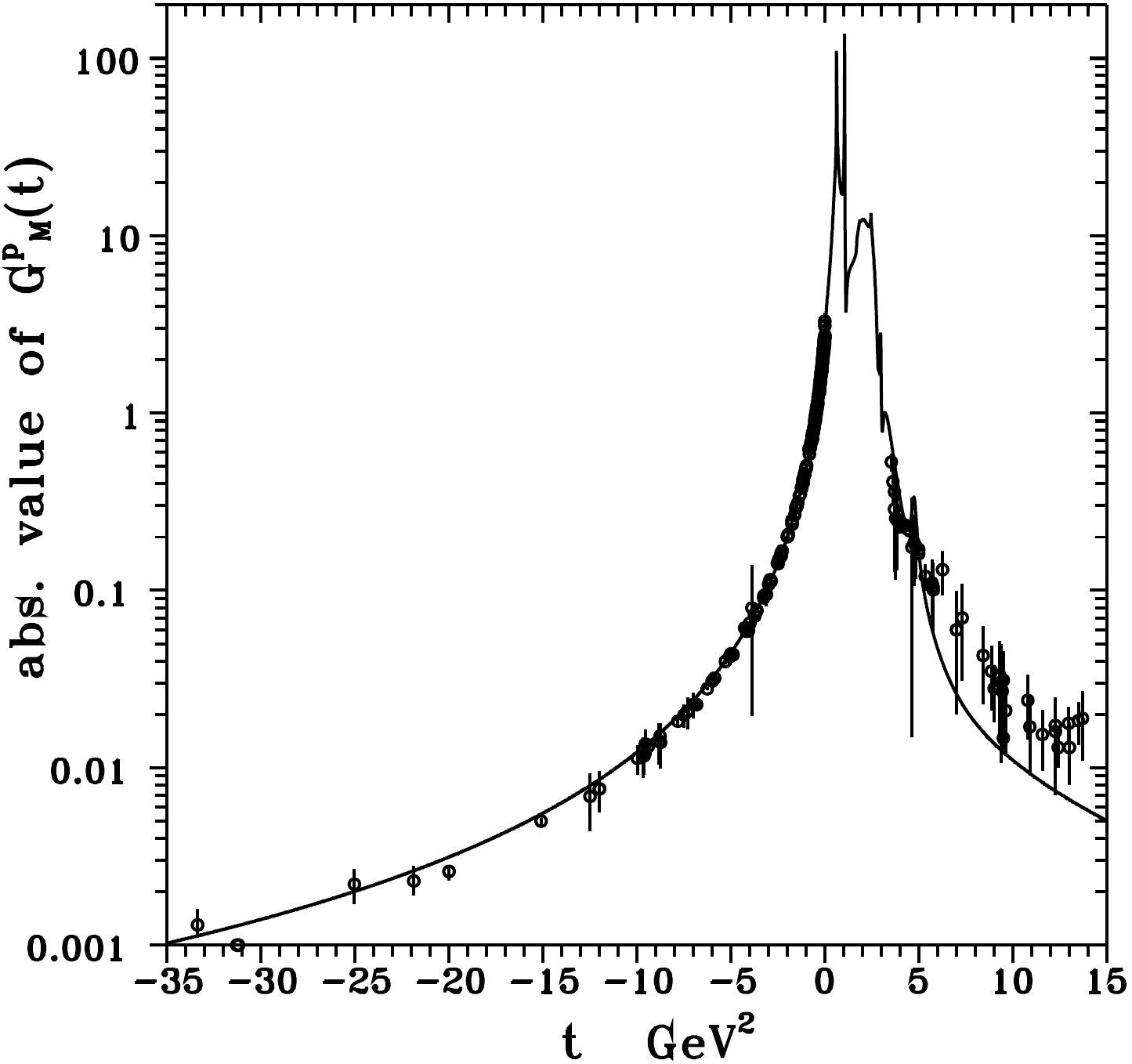}\\
\caption{The proton EM structure $U\&A$ model given by (\ref{FN1s})-(\ref{FN2v}) and (\ref{pEMFFs}) with numerical values of parameters of TABLE I reproduce
existing data quite well.\label{pemstth}}
\end{figure}

   Behaviors of $G^p_E(t)$ and $G^p_M(t)$ with numerical values of the parameters of TABLE I and their comparison with existing data is presented in Fig. \ref{pemstth}.
\begin{table}
\centering
\caption{Results of analysis of the proton EM FFs data with $\chi^2/ndf=1.74$.}
\begin{tabular}{ll}
\hline\hline
  $t^{1s}_{in}= (1.6750\pm 0.0363) $GeV$^2$ & $t^{1v}_{in}= (2.9683\pm 0.0091) $GeV$^2$\\
  $t^{2s}_{in}= (1.8590\pm 0.0023) $GeV$^2$ & $t^{2v}_{in}= (2.4425\pm 0.0208) $GeV$^2$\\
  $(f^{(1)}_{\omega' NN}/f_{\omega'})= -0.2937\pm 0.0015$ \quad& $(f^{(1)}_{\phi' NN}/f_{\phi'})= -0.5298\pm 0.0027$\\
  $(f^{(1)}_{\omega NN}/f_{\omega})= 0.6384\pm 0.0025$ & $(f^{(1)}_{\phi NN}/f_{\phi})= -0.0271\pm 0.0005$\\
  $(f^{(2)}_{\phi' NN}/f_{\phi'})= 0.3075\pm 0.0156$ & $(f^{(2)}_{\omega NN}/f_{\omega})= 0.1676\pm 0.0377$\\
  $(f^{(2)}_{\phi NN}/f_{\phi})= 0.1226\pm 0.0035$ & $(f^{(1)}_{\rho NN}/f_{\rho})= -0.0802\pm 0.0014$\\
\hline\hline
\end{tabular}
\end{table}

\section{Theoretical predictions for neutron electromagnetic form factors behaviors and inequality $\sigma_{tot}(e^+e^- \to n \bar n) > \sigma_{tot}(e^+e^- \to p \bar p)$}

\begin{figure}
    \includegraphics[width=0.35\textwidth]{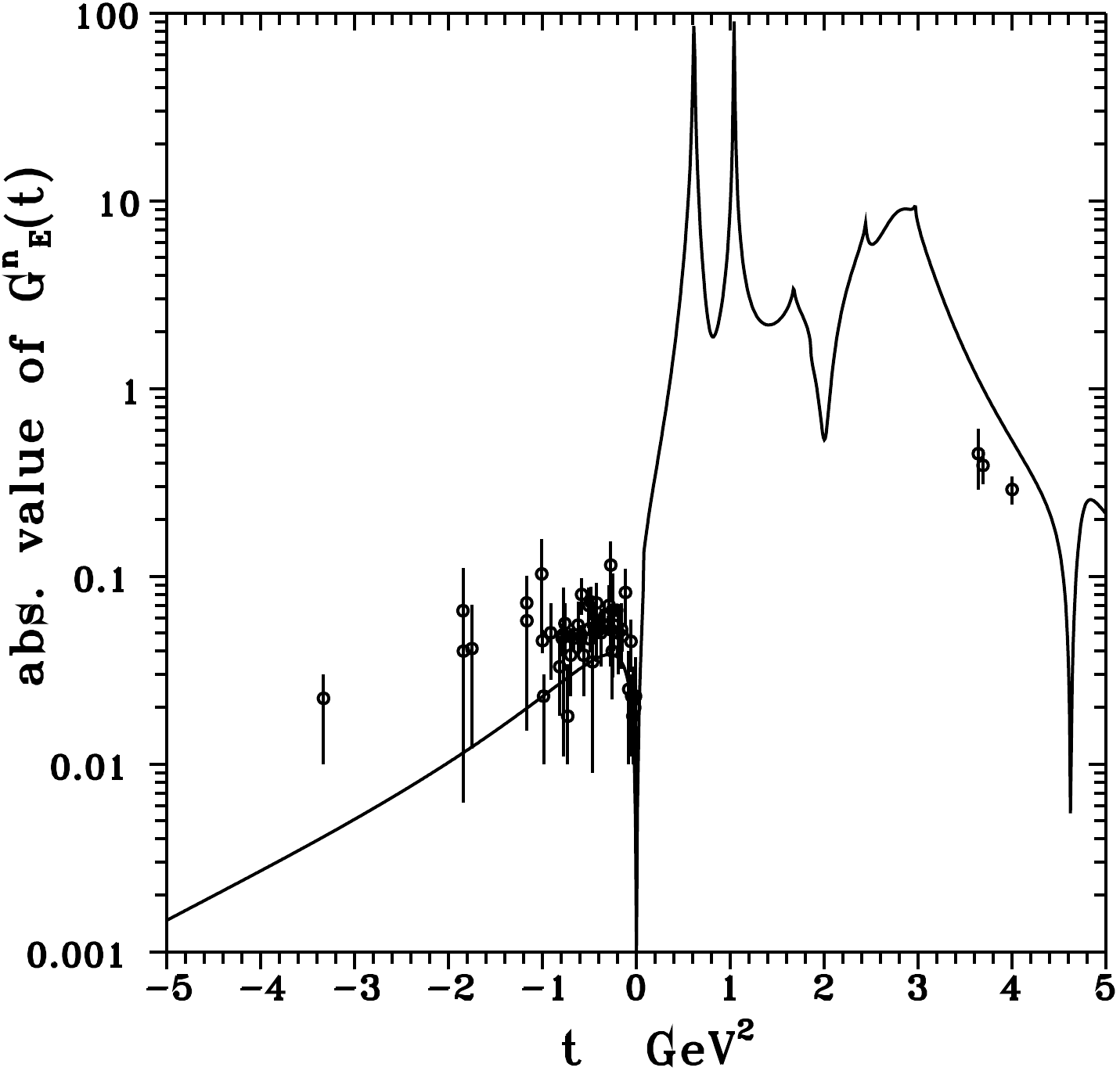}\hspace{0.3cm}
    \includegraphics[width=0.35\textwidth]{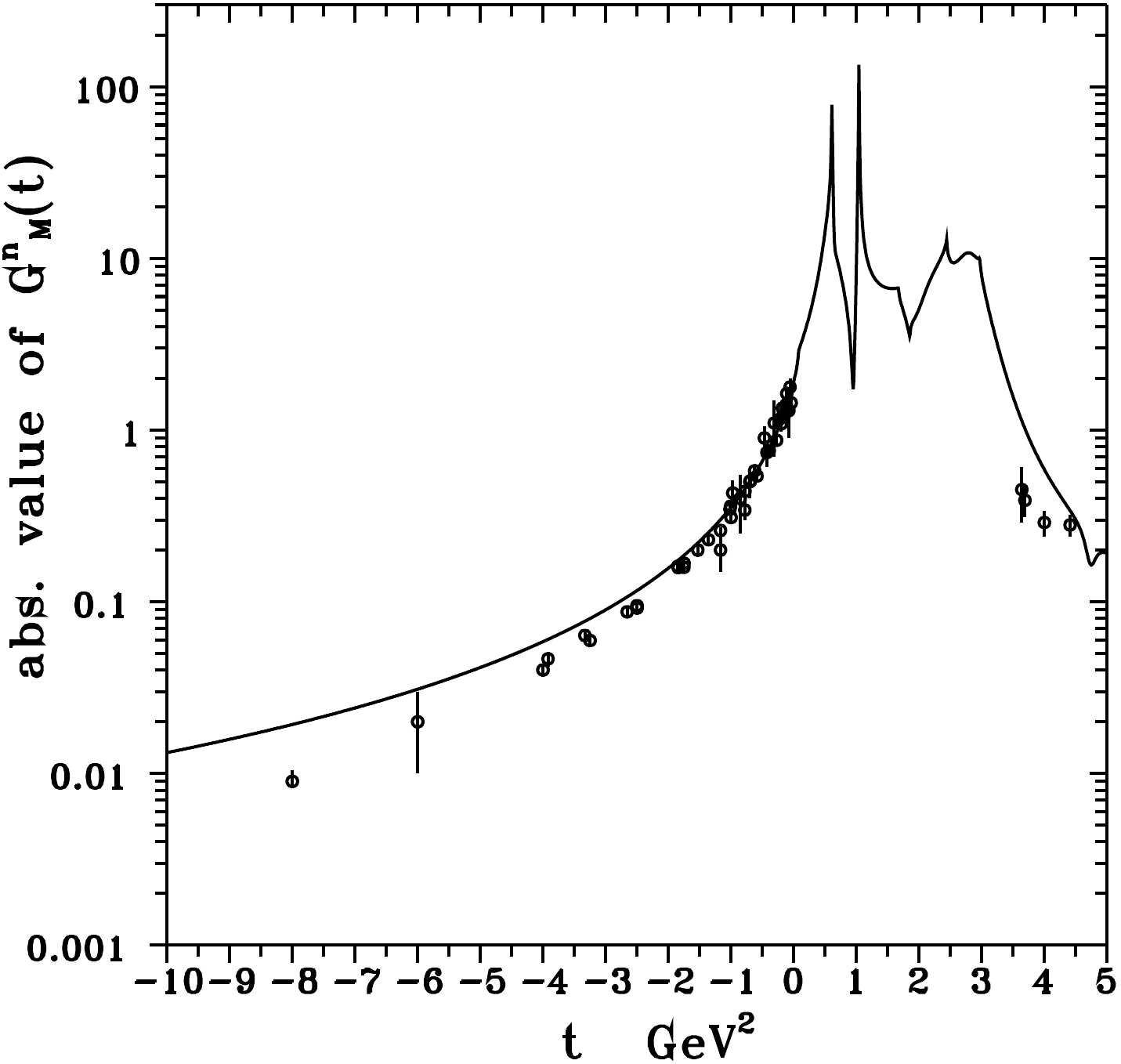}\\
\caption{Theoretical prediction of the neutron EM FFs behaviors by the neutron EM structure $U\&A$ model given by (\ref{FN1s})-(\ref{FN2v}) and (\ref{nEMFFs}) with the numerical values of parameters from TABLE I. The neutron experimental points are presented for a comparison with predicted behaviors.\label{nemstth}}
\end{figure}

   The experimental information on the neutron EM FFs is very poor. In the spacelike region due to an instability of the neutron, which prevents a preparation of
a stable neutron target for experimental investigation of the elastic scattering of electrons on neutrons and instead only some indirect methods of measurements of the neutron EM FFs with light nuclei could be carried out. On the other hand in the timelike region due to a problem with a detection of unstable neutral neutrons in the annihilation process $e^+e^- \to n \bar n$. As a result the use of rare and unreliable neutron data \cite{hanson}-\cite{riordan} for a numerical evaluation of the free parameters of the neutron EM structure $U\&A$ model given by (\ref{FN1s})-(\ref{FN2v}) and (\ref{nEMFFs}) led to a completely distinct values from those given in TABLE I, though on the base of the relations (\ref{DPp}) and (\ref{DPn}) they have to be identical.

   Therefore we have substituted the numerical values of free parameters of TABLE I, to be determined in the analysis of only proton data, into relations
$G^n_E(t), G^n_M(t)$ (\ref{nEMFFs}) for neutron EM FFs and by this way the neutron EM structure $U\&A$ model given by (\ref{FN1s})-(\ref{FN2v}) and (\ref{nEMFFs}) is
completely determined. Predicted behaviors of the neutron EM FFs have been achieved by means of such model pure theoretically as it is exhibited in Fig. \ref{nemstth}, without an utilization of any experimental point shown at the same figure for a comparison. The correctness of $G^n_E(t), G^n_M(t)$ behaviors is confirmed not only by coincidence of the theoretically calculated curves with deficient neutron EM FFs data, but also by roughly evaluated neutron EM radii values, $<r^2>_{nE}=-0.125$ fm$^2$ and $\sqrt{<r^2>}_{nM}=0.901$ fm, to be compared with PDG \cite{PDG} neutron radii, $<r^2>_{nE}=-0.1161\pm0.0022$ fm$^2$ and $\sqrt{<r^2>}_{nM}=0.864\pm0.009$ fm, respectively.

\begin{figure}
    \includegraphics[width=0.35\textwidth]{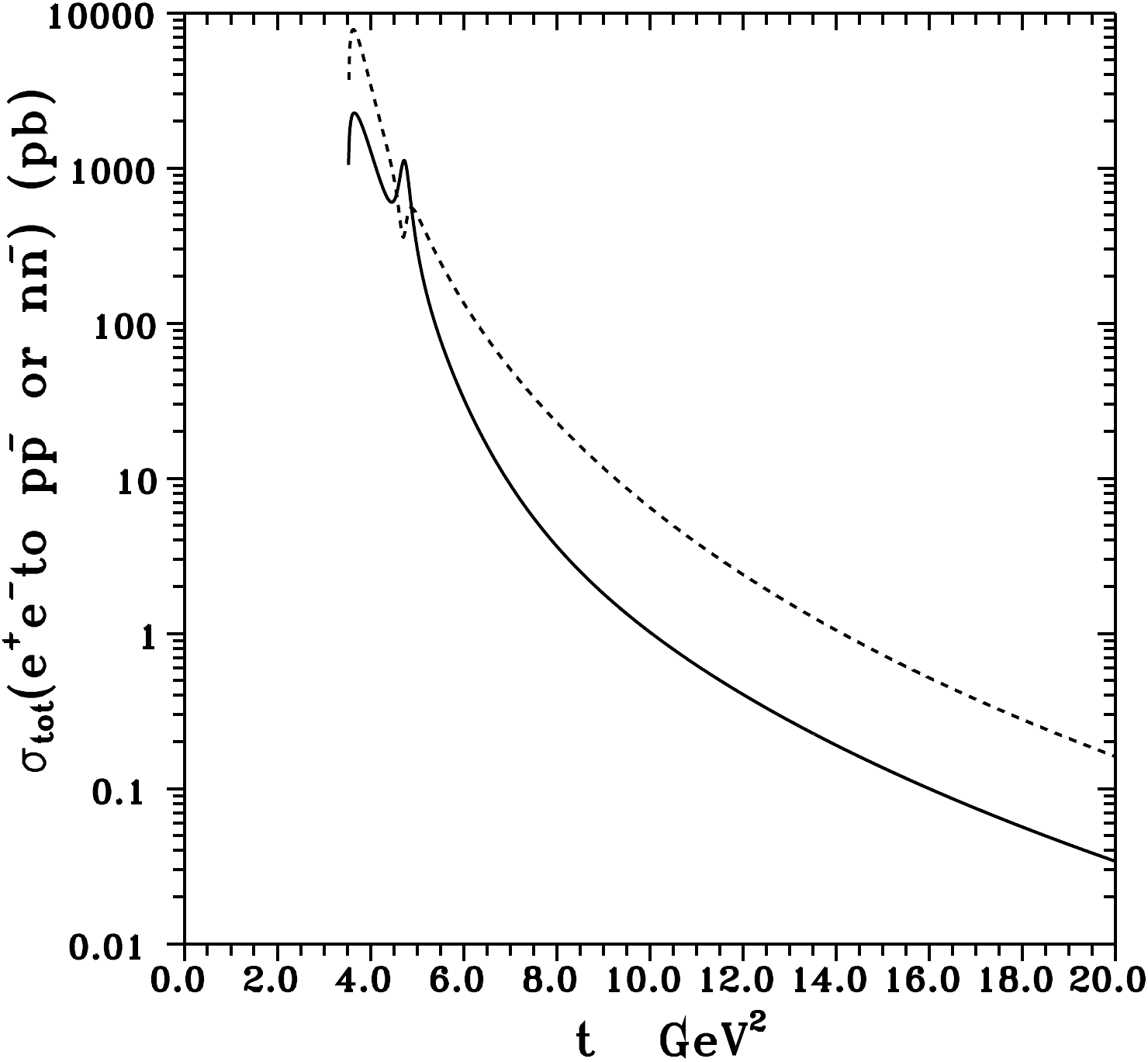}
\caption{Theoretically predicted behavior of the neutron total cross section $\sigma_{tot}(e^+e^- \to n \bar n)$ (dashed line) and its comparison with the proton total cross section $\sigma_{tot}(e^+e^- \to p \bar p)$ (full line) based on the parameters of TABLE I.\label{pncsec}}
\end{figure}

   So, by an employment of only the existing proton EM FFs data, the earlier constructed nucleon EM structure $U\&A$ models to be based on the analyticity and the mixed
transformation properties of the EM current $J^{EM}_\mu (0)$ under the rotation in the isospin space, the correct behaviors of both, the proton EM FFs as presented in Fig. \ref{pemstth} and theoretically predicted neutron EM FFs as presented in Fig. \ref{nemstth} are now available. Substituting these behaviors of EM FFs into expressions for the corresponding total cross sections $\sigma_{tot}(e^+e^- \to p \bar p)$ given by (\ref{pEMFFs}) and $\sigma_{tot}(e^+e^- \to n \bar n)$ given by (\ref{nEMFFs}), one obtains their explicit behaviors and by a comparison of both curves in one Fig. \ref{pncsec}, the following inequality $\sigma_{tot}(e^+e^- \to n \bar n) > \sigma_{tot}(e^+e^- \to p \bar p)$, is definitely confirmed, which is not physically comprehensible till now.

\section{Conclusions}

   The absolute values of the neutron EM FFs, $G^n_E(t), G^n_M(t)$, are predicted theoretically in the whole region of their definition
without exploiting of any experimental point on them. This could be carried out by the fact that the proton EM structure $U\&A$ model given by (\ref{FN1s})-(\ref{FN2v}) and (\ref{pEMFFs}) and the neutron EM structure $U\&A$ model given by (\ref{FN1s})-(\ref{FN2v}) and (\ref{nEMFFs}) are compound of the same flavor-independent isoscalar $F^N_{1s}(t), F^N_{2s}(t)$ and isovector $F^N_{1v}(t), F^N_{2v}(t)$ parts of the corresponding Dirac and Pauli FFs and therefore they depend on the same free parameters. They have been evaluated numerically, see TABLE I, in the analysis of only more reliable proton EM FFs data.

   On the conclusion we would like to emphasize that even if more compact and precise proton EM FFs data will appear in as wide region of their definition as
possible, then more accurate predictions for the absolute values of the neutron EM FFs will be achieved by the approach discussed in this paper and more convincing
results for the  inequality $\sigma_{tot}(e^+e^- \to n \bar n) > \sigma_{tot}(e^+e^- \to p \bar p)$ between the total cross sections under consideration, especially near to the nucleon-antinucleon threshold, can be found.

   The above wish has been already started to be realistic with the first time in the world BESS III Collaboration measurements \cite{Ab1} of the proton angular
distribution in the process $e^+e^- \to p \bar p$ and a simultaneous determination of both proton EM FFs in timelike region with the unprecedented accuracy. Nevertheless, such results should be verified by some alternative method of FFs measurements, e.g. by a determination of the proton EM FFs behaviors in timelike region by a measurement of vector and tensor polarizations in the $e^+e^- \to p \bar p$ process \cite{DDR}. There is already appeared the case in the past when polarization experiments in the $e^-p \to e^-p$ process \cite{jones}-\cite{pucket} refuted a 50-year belief in a dipole behavior of the proton electric FF from the normalization point to $-\infty$, indicated by the spacelike data obtained by the Rosenbluth method.

\medskip

   The authors would like to thank Eric Bartos for valuable discussions.
   The support of the Slovak Grant Agency for Sciences VEGA, grant No.2/0153/17, is acknowledged.

\end{document}